\documentclass[aps,prl,reprint,superscriptaddress,showpacs]{revtex4-1}
\usepackage{amssymb}

\usepackage{graphicx}
\usepackage{amsmath}

\begin{document}

\title{Quantization of entropy in a quasi-two-dimensional electron gas}
\author{A. A. Varlamov}

\affiliation{CNR-SPIN, Viale del Politecnico 1, I-00133, Rome, Italy}
\author{A. V.~Kavokin}
\affiliation{CNR-SPIN, Viale del Politecnico 1, I-00133, Rome, Italy and Physics
	and Astronomy School, University of Southampton, Southampton, SO171BJ, UK}
\author{Y.~M. Galperin}
\affiliation{Department of Physics, Oslo University, P. O. Box 1048,
	Blindern, 0316 Oslo, Norway} 
\affiliation{A. F. Ioffe Physico-Technical Institute RAS, 194021 St.
	Petersburg, Russian Federation}
\date{\today }

\begin{abstract}
We demonstrate that the partial entropy of a two-dimensional electron gas
(2DEG) exhibits quantized peaks at resonances between the chemical potential
and electron levels of size quantization. In the limit of no scattering, the
peaks depend only on the subband quantization number and are independent on
material parameters, shape of the confining potential, electron effective
mass and temperature. The quantization of partial entropy is a signature of
a topological phase transition in a 2DEG. In the presence of stationary
disorder, the magnitude of peaks decreases. Its deviation from the quantized
values is a direct measure of the disorder induced smearing of the
electronic density of states.
\end{abstract}

\maketitle

\paragraph{Introduction. --}

Low-dimensional electronic devices are important building blocks for quantum
electronics. This is one of the reasons of great interest to these systems.
Another reason is that size quantization of the electronic states in
low-dimensional systems leads to quantization of their thermal and transport
properties. The most famous are the integer~\cite{vonKlitzing} and
fractional~\cite{Tsui82} quantum Hall effect in two-dimensional electron gas
(2DEG) and conductance quantization of quasi-one-dimensional channels~\cite%
{Wharam,*VanWees}.

In this Letter, we address the major thermodynamic quantity -- entropy -- of
a quasi-two-dimensional electron gas. An elegant way to measure directly the
entropy per electron, $s\equiv (\partial S/\partial n)_T$, was recently
demonstrated~\cite{Pudalov}. Here $T$ is temperature. We will show that the
quantization of the energy spectrum of quasi- 2DEG into sub-bands leads to a
very specific quantization of the entropy: $s$ exhibits sharp maxima as the
chemical potential $\mu $ passes through the bottoms of size quantization
subbands ($E_{i}$). The value of the entropy in the $N$-th maximum depends 
\textit{only} on the number of the maximum, $N$: 
\begin{equation}  \label{s}
s\vert_{\mu = E_n} \equiv \left(\frac{\partial S}{\partial n}\right)_{T,\,
\mu = E_n} = \frac{\ln 2}{N - 1/2}\, .
\end{equation}
In the absence of scattering this result is independent of the shape of the
transversal potential that confines 2DEG and of the material parameters
including the electron effective mass and dielectric constant.

The universality of the above quantization rule can be broken both by
disorder and by electron-electron interactions. Using a simple model of
Lorentzian smearing of the electronic spectrum we show that it leads to the
relative correction of $\sim \hbar /T\tau $, where $\tau $ is the electron
life time. For the case of a single-band 2DEG the role of the
electron-electron interaction in the partial entropy was investigated in~%
\cite{Pudalov}, see references therein for a review. We believe that Eq.~%
\eqref{s} can be used as a benchmark allowing to judge on the importance of
the disorder and interactions. In this Letter we report on the analytical
dependence of the partial entropy on the chemical potential accounting for
the smeared density of states of 2DEG at the electron quantization levels.
We reveal the quantization of partial entropy at resonances of the chemical
potential and electron quantization levels and discuss the accuracy of the
obtained expression for the quantized entropy in the presence of disorder
and electron- electron interactions.

\paragraph{General expressions. --}

In the absence of scattering, the density of electronic states (DOS) in a
non-interacting 2DEG has a staircase-like shape~\cite{Ando}, 
\begin{equation}
g(\mu )=\frac{m^{\ast }}{\pi \hbar ^{2}}\sum\limits_{i=1}^{\infty }\theta
\left( \mu -E_{i}\right) ,  \label{nuclean}
\end{equation}%
with $m^{\ast }$ being the electron effective mass and $\theta \left(
x\right) $ being the Heaviside theta-function. Elastic scattering of
electrons on defects and impurities that is necessarily present in realistic
systems, leads to the smearing of the steps of the density of states. A
simple way to account for this smearing is to introduce a finite life-time, 
$\tau $, of an electron. That results in the replacement of a Dirac delta-function by a Lorentzian in the derivative of the density of states: $%
\theta ^{\prime }(E)=\delta (E)\rightarrow \hbar \tau ^{-1}/\pi (E^{2}+\hbar
^{2}\tau ^{-2})$. Integration of the latter expression leads to the
replacement $\theta (E)\rightarrow \tilde{\theta}(E),$ where 
\begin{equation}
\tilde{\theta}\left( E\right) =\frac{1}{2}+\frac{1}{\pi }\arctan \left( 
\frac{E\tau }{\hbar }\right) .  \label{theta}
\end{equation}%
We will focus on a case where $T\gg \hbar /\tau $, that corresponds to a
relatively clean sample. At the same time, temperatures are supposed to be
not too high, $T\ll \Delta _{Nj}=|E_{N}-E_{j}|,\forall j\neq N$. In
addition, we assume that the transport is adiabatic~\cite{Beenakker}, i.e.
there are no elastic inter-band transitions due to backscattering.

To find the partial entropy, $s$, we use the Maxwell relation 
\begin{equation}
s=\left( \frac{\partial S}{\partial n}\right) _{T}=-\left( \frac{\partial
\mu }{\partial T}\right) _{n}=\left( \frac{\partial n}{\partial T}\right)
_{\mu }\left( \frac{\partial n}{\partial \mu }\right) _{T}^{-1}.
\label{fullderiv}
\end{equation}%
The relationship between the electron concentration $n$, chemical potential $%
\mu $ and temperature $T$ can be found integrating Eq. (\ref{nuclean}) over
energies with the Fermi-Dirac distribution and accounting for the
renormalization (\ref{theta}): 
\begin{equation}
n\left( \mu ,T\right) =\frac{m^{\ast }}{\pi \hbar ^{2}}\sum\limits_{i=1}^{%
\infty }\int\limits_{0}^{+\infty }\frac{\tilde{\theta}(E-E_{i})}{\exp \left( 
\frac{E-\mu }{T}\right) +1}\,dE.  \label{ngen}
\end{equation}%
Calculating the partial derivatives of the electron concentration over
temperature and chemical potential one can express them in the form of sums
over the subband levels averaged over energy with temperature and impurities
smearing factors: 
\begin{eqnarray}
\left( \frac{\partial n}{\partial T}\right) _{\mu } &=&\frac{m^{\ast }}{\pi
^{2}\hbar ^{2}}\!\sum\limits_{j=1}^{\infty }\int\limits_{-\infty }^{+\infty }%
\frac{f_{n}(z)}{\cosh ^{2}z}dz,  \label{dndT} \\
\left( \frac{\partial n}{\partial \mu }\right) _{T} &=&\frac{m^{\ast }}{2\pi
^{2}\hbar ^{2}}\sum\limits_{j=1}^{\infty }\int\limits_{-\infty }^{+\infty }%
\frac{f_{\mu }(z)}{\cosh ^{2}z}dz,  \label{dndm} \\
f_{n}(z) &\equiv &z\arctan \left[ \left( 2Tz+\!\delta _{N}+\Delta
_{Nj}\right) \tau \right] ,  \notag \\
f_{\mu }(z) &\equiv &\pi /2+\arctan \left[ \left( 2Tz+\!\delta _{N}+\Delta
_{Nj}\right) \tau \right] .  \notag
\end{eqnarray}%
Here $\delta _{N}=\mu -E_{N}$ (we assume $|\delta _{N}|\ll \Delta _{NN\pm1}$).
We have also taken into account that \ $\mu \rightarrow E_{N}\gg T$ and
extended the lower limit of integration up to $-\infty .$

\paragraph{Non-interacting 2DEG in the absence of disorder. --}

We start the analysis with the case of a clean material, where one can
neglect the smearing of electron states and replace $\arctan z\rightarrow
(\pi /2)\mathop{\mathrm{sign}}z$. In this case the principal contribution to
the derivative (\ref{dndT}) gives the level closest to the the chemical
potential: 
\begin{equation*}
\left( \frac{\partial n}{\partial T}\right) _{\mu \rightarrow E_{N}}=\frac{%
m^{\ast }}{\pi \hbar ^{2}}\left[ \ln \left( 2\cosh \frac{\delta _{N}}{2T}%
\right) -\frac{\delta _{N}}{2T}\tanh \frac{\delta _{N}}{2T}\right] .
\end{equation*}%
The contributions of other levels are exponentially small, they turn to be
of the order of  $\exp (-\Delta _{N,N\pm 1}/T)$. In Eq.~(\eqref{dndm}), the
lowest $N-1$ levels provide the same universal, independent on chemical
potential and temperature, contributions, while the shape of the line is
determined by the N-th level. We obtain: 
\begin{equation}
\left( \frac{\partial n}{\partial \mu }\right) _{\mu \rightarrow E_{N}}=%
\frac{m^{\ast }}{\pi \hbar ^{2}}\left( N-\frac{1}{2}\right) +\frac{m^{\ast }%
}{2\pi \hbar ^{2}}\tanh \frac{\delta _{N}}{2T}.  \notag
\end{equation}%
The contributions of the higher levels ($j>N$) are exponentially small.

\begin{figure}[th]
\includegraphics[width=0.7\columnwidth]{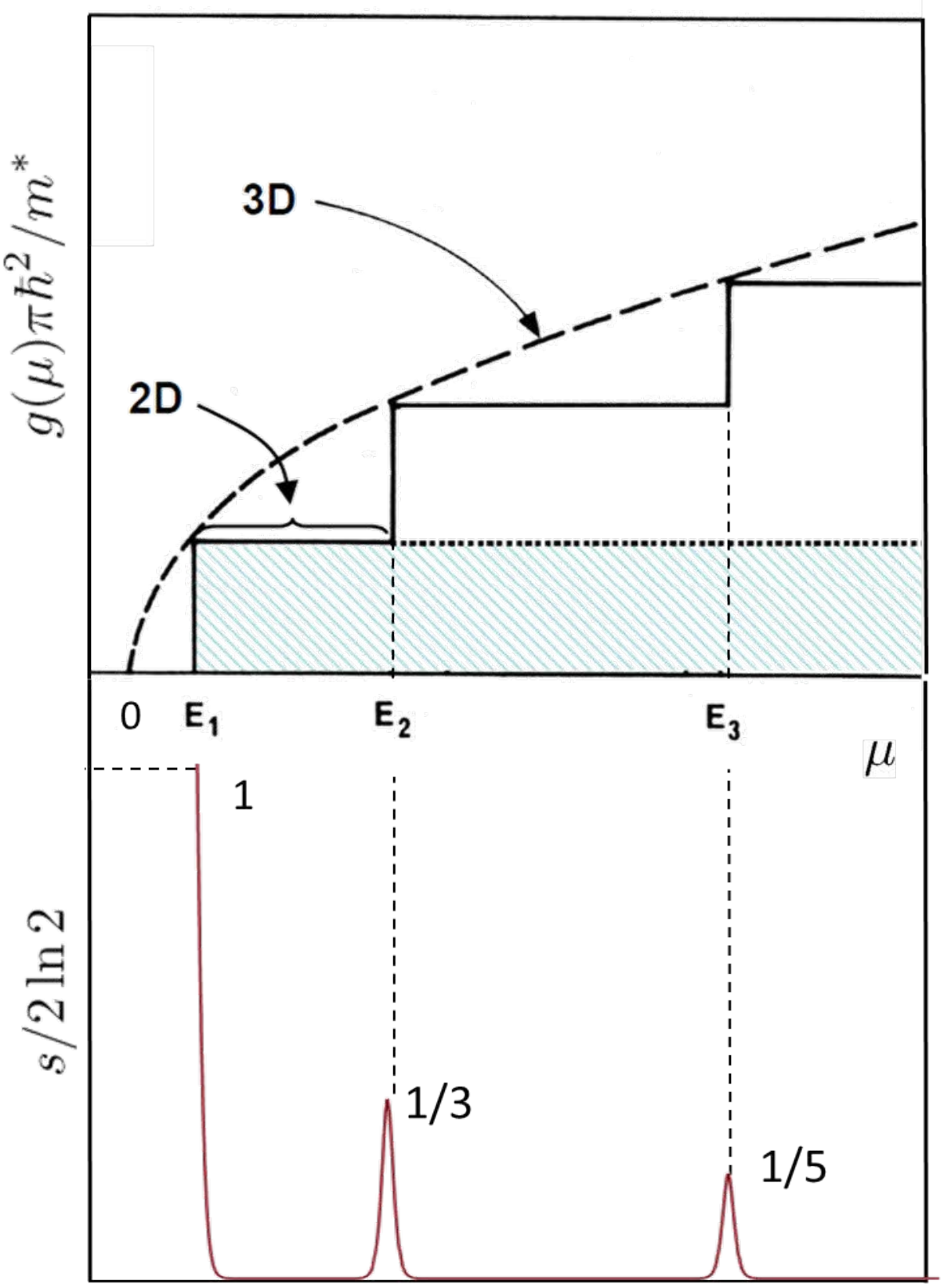}
\caption{Schematic representation of the dependencies of the electronic
density of states (upper panel) and the partial entropy (lower panel) as
functions of the chemical potential.}
\label{fig1}
\end{figure}

Finally, the expression for the partial entropy Eq. (\ref{fullderiv}), valid for any spectrum of size quantization 
$E_{j},$\ takes the form: 
\begin{eqnarray}  \label{fin}
s_{\mu \rightarrow E_{N}} &=&\frac{\ln \left( 2\cosh \frac{\delta _{N}}{2T}%
\right) -\frac{\delta _{N}}{2T}\tanh \frac{\delta _{N}}{2T}}{\left(
N-1/2\right) +\frac{1}{2}\tanh \frac{\delta _{N}}{2T}}  \label{final} \\
&=&\left\{ 
\begin{array}{cc}
\frac{|\delta _{N}|}{T}\frac{\exp \left( -\frac{|\delta _{N}|}{T}\right) }{%
N-1+\exp \left( -\frac{|\delta _{N}|}{T}\right) }, & \delta _{N}\ll -T, \\ 
\frac{\ln 2}{N-1/2}, & 0<\delta _{N}\ll T, \\ 
\frac{\delta _{N}}{TN}\exp (-\delta _{N}/T), & \delta _{N}\gg T.%
\end{array}%
\right.   \label{finas}
\end{eqnarray}%
This expression predicts the existence of quantized peaks of the partial
entropy $s$ at $\mu =E_{N}$, their magnitudes being dependent only on the
subband number. The dependence of $s$ on the chemical potential is
schematically shown in Fig.~\ref{fig1}, lower panel. The quantized peaks of
the partial entropy correspond to the steps of the density of states shown
in the upper panel of the same figure. 
The shape of the peaks in Eq. (\ref{final}) is asymmetric, that corresponds
to the step-like changes in the density of electronic states as a function
of the chemical potential of the 2DEG.

\paragraph{Effect of scattering. --}

Equations (\ref{dndT}) and (\ref{dndm}) allow one to estimate the effect of
scattering on the heights of the peaks. A straightforward analysis shows
that the contribution of the lower subbands is of the order of $%
\sum\limits_{j=1,j\neq N}^{\infty }\mathcal{O}\left[ \left( \Delta _{Nj}\tau
/\hbar \right) ^{-n}\right] $. Here $n=1$ for $(\partial n/\partial \mu )_{T}
$ and $n=2$ for $(\partial n/\partial T)_{\mu }$. These sums are cut off at $%
j_{\max }\sim \hbar /T\tau .$ For the equidistant spectrum (parabolic
potential), or $E_{N}\sim N^{2/3}$ (eigenvalues of the Airy functions, in
the case of the linear potential) they give small contributions of the order 
$\left( \Delta \tau /\hbar \right) ^{-n}\left( T/\Delta \right) ^{n-1}$($%
\Delta $ is the characteristics scale of inter-level distances).

The contribution of the $N$-th subband to $(\partial n/\partial \mu)_T$ can
be calculated exactly  leading to the replacement  
\begin{equation*}
\tanh \frac{\delta _{N}}{2T} \to \mathop{\mathrm{Re}}\left[\tanh \frac{
\delta _{N}}{2T} -i\frac{\hbar}{2\tau T}\right] 
\end{equation*}
in Eq. (\ref{final}), i.e. to appearance of the corrections of the order $%
\mathcal{O}\left[ \left( \hbar /T\tau \right)^{2}\right] $. Yet, the
dominant effect of impurities is due to $(\partial n/\partial T)_\mu$. The
asymptotic analysis of Eq. (\ref{dndT}) shows that the magnitude of the peak
in $s$ is suppressed by the elastic scattering of electrons as%
\begin{equation*}
s_{\mu =E_{N}}=\frac{\ln 2-\left(\hbar / \pi T\tau \right)}{N-1/2}.
\end{equation*}%
This simple relation allows to characterize the degree of disorder in a 2DEG.

\paragraph{Discussion. --}

The dependence of the partial entropy $s$ on the chemical potential can be
interpreted in the following way. At low temperatures, the main contribution
to the entropy is provided by the electrons having energies in the vicinity
of the Fermi level, the width of the `active' layer being $\sim T$. If the
electron DOS is constant within the layer then by adding an electron one
does not change the entropy. Hence, the entropy is independent of the
chemical potential, $(\partial S/\partial n)_{T}\rightarrow 0$. However, if
the bottom of one of the subbands falls into the active layer, the number of
`active' states becomes strongly dependent on the chemical potential. In
this case, adding an electron to the system, one changes the number of
'active' states in the vicinity of the Fermi surface. Consequently, the
partial entropy strongly increases. The peaks of the partial entropy
correspond to the resonances of the chemical potential and electron size
quantization levels. The further increase of the chemical potential brings
the system to the region of the constant density of states, where the
partial entropy vanishes again.

The intersections by the chemical potential of the levels of electron size
quantization, $\delta _{N}=0$, can be considered as the points of
topological phase transitions in a 2DEG, where the Fermi surface acquires a
new component of topological connectivity. Corresponding anomalies in the
thermodynamic and transport characteristics, in particular, thermoelectric
coefficient related to the peculiarities of the energy dependence of the
electron momentum relaxation time have been studied experimentally and
theoretically in Refs.~\cite{Z1,Z2} and \cite{BPV92}, respectively. Here we
present an analytical theory of purely thermodynamic anomalies.

In the asymptotic expression \eqref{finas} for strongly negative $\delta_N $
(but $|\delta_N| \gg T$) , the item $\exp(-|\delta_N|/T)$ in the denominator
can be neglected for all $N>1$. However, it becomes important for $N=1$.
Indeed, at $\mu < E_1$ the partial entropy increases as $|\mu -E_1|/T$ with
decreasing $\mu$. This is a manifestation of the crossover from the Fermi
distribution to the Boltzmann one when the chemical potential falls into the
gap in the spectrum. The region $\mu < E_1$ is not shown in Fig.~\ref{fig1}
in order to keep the peaks for $N=2,3$ visible.

At $T\rightarrow 0$ (yet $T\gtrsim \hbar /\tau $ ) the peaks of $s$ are
located at $\mu \rightarrow E_{N}$, $N>1$, the maximal values being $s_{\max
}(N)=\ln 2/(N-1/2)$. At finite $T$ the peaks acquire finite widths of the
order of $T$ and shift toward negative values of $\mu -E_{N}$. 
\begin{figure}[th]
\centerline{
\includegraphics[width=.7\columnwidth]{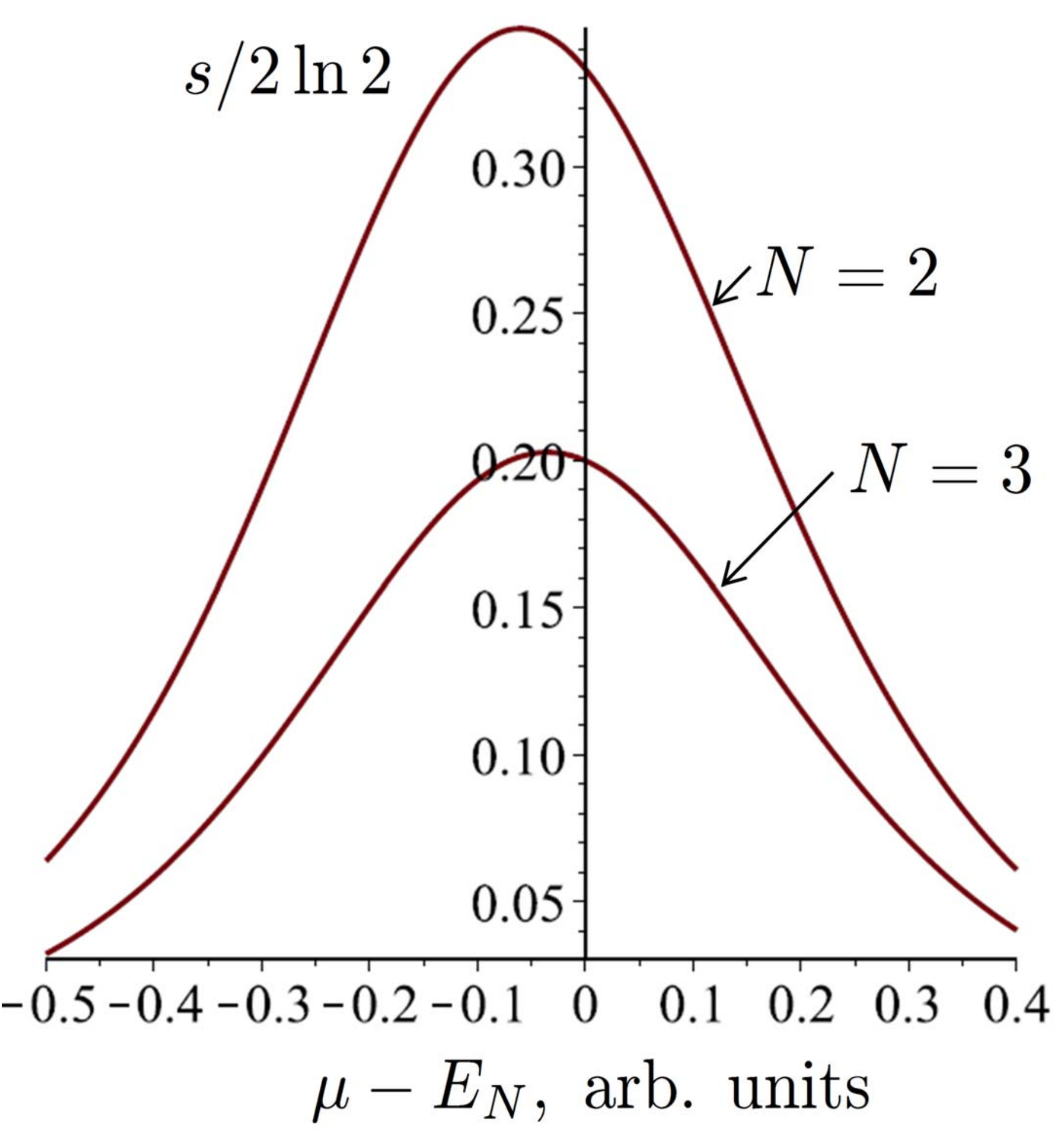}
}
\caption{Dependence $s(\protect\mu -E_{N})$ for $N=2,3$. }
\label{fig2}
\end{figure}
The peaks of $s$ for $N=2$ and 3 are shown in Fig.~\ref{fig2}, their
characteristics are given in Table~\ref{tab}. The reason of the peaks'
asymmetry is the difference in partial densities of states above and below
the chemical potential. The relative difference between the DOSs decreases
with increase of $N$, therefore the peaks become more symmetric. 
\begin{table}[h]
\begin{tabular}{|c|c|c|c|}
\hline
N & $\delta_{\max}/2T$ & $s_{\max}/2\ln 2$ & $s\vert_{\mu = E_N}/2\ln 2$ \\ 
\hline
2 & -0.24 & 0.347 & 1/3 \\ \hline
3 & -0.14 & 0.203 & 1/5 \\ \hline
4 & -0.01 & 0.144 & 1/7 \\ \hline
\end{tabular}%
\caption{Peaks in the partial entropy. }
\label{tab}
\end{table}

Interestingly, our result for $(\partial \mu/\partial T)_n \equiv -s$ at $%
\mu = E_1$ and $\tau \to \infty$ coincides with the expression for the same
derivative obtained in Refs.~\cite{Bob71,Abrikosov} for a two-dimensional
superconductor. The quantized dip in this derivative is associated with the
step in the density of electronic states which changes from zero inside the
superconducting gap to $m^{\ast }/\pi \hbar ^{2}$ above the gap. A
remarkable fact is that the value of the effective mass $m^{\ast }$ does not
enter the result.

Note that the variation of the chemical potential as a function of
temperature can be measured by resonant optical transmission spectroscopy of
the fundamental absorption edge in modulation doped semiconductor quantum
wells, see, e.g.,~\cite{Dutton1958}.

Now let us briefly discuss the role of electron-electron (e-e) interactions,
which are neglected in the above formalism. E-e interactions become
noticeable for the electronic states sufficiently close to the subbands'
bottoms. Characterizing the importance of e-e interaction by the parameter $%
r_{s}$~\cite{Mahan} we conclude that $r_{s}\approx 1$ for the upper filled
subband at 
\begin{equation*}
\sqrt{\frac{2\delta _{N}}{m^{\ast }}}\approx \frac{e^{2}}{\kappa \hbar }%
\rightarrow \delta _{N}\approx \frac{m^{\ast }}{2}\left( \frac{e^{2}}{\kappa
\hbar }\right) ^{2},
\end{equation*}%
with $\kappa $ being a dielectric constant. \ Putting $m^{\ast }=0.1m_{0}$ and $\kappa =10$ we
get $\delta _{N}\gtrsim 2\times 10^{-14}$~erg. If $\delta _{N}$ is less than
this value one can expect a Fermi-liquid renormalization of the electron
spectrum, in particular, of the effective mass $m^{\ast }$. Fortunately, $%
m^{\ast }$ doesn't enter the expressions for the peaks of $s$. However,
additional correction proportional to $(\partial m^{\ast }/\partial n)$ can
appear in the expression \eqref{dndm} for the thermodynamic DOS. These
corrections for $N=1$ are discussed in Ref.~\onlinecite{Pudalov} and
references therein. Another possibility of evidencing the interaction
effects is establishing a special regime of a correlated 2D charged plasma~%
\cite{Novikov} also explored in Ref.~\onlinecite{Pudalov}. Here we do not
consider this particular case. In general, comparing the experimental
results with the universal expression obtaied here~\eqref{fin} allows one to
judge on the role of electron-electron correlations in a system under study.

In conclusion, we have analytically derived the partial entropy $(\partial
S/\partial n)_T$ in a non-interacting quasi-2DEG in the vicinity of electron quantization levels. The values of the peaks in $(\partial S/\partial n)_T$
appear to be independent of the effective mass, dielectric constant and
other material parameters. The partial entropy may be directly measured
either by temperature modulation~\cite{Pudalov}, or by the optical
transmission measurements~\cite{Dutton1958}.

AV and AK acknowledge partial support from the HORISON 2020 RISE project CoExAn (grant agreement 644076). YG acknowledges 
partial support from the EU FP7 project INFERNOS (grant agreement 308850).


\begin{thebibliography}{99}
\bibitem{vonKlitzing} K. von Klitzing and G. Dorda, and M. Pepper, \prl~%
\textbf{45}, 494 (1980).

\bibitem{Tsui82} D. C. Tsui, H. L. Stormer, A. C. Gossard, \prl~\textbf{48},
1559 (1982).

\bibitem{Wharam} D. A. Wharam, T. J. Thornton, R. Newbury, M. Pepper, H.
Ahmed, J. E. F. Frost, D. G. Hasko, D. C. Peacock, D. A. Ritchie and G. A.
C. Jones. J. Phys. C \textbf{21}, L209 (1988).

\bibitem{VanWees} B. J. van Wees, H. van Houten, C. W. J. Beenakker, J. G.
Williamson, L. P. Kouwenhoven, D. van der Marel, and C. T. Foxon \prl~%
\textbf{60}, 848 (1988).

\bibitem{Pudalov} A. Yu. Kuntsevich, V. M. Pudalov, I. V. Tupikov, I. S.
Burmistrov,  Nature Communications \textbf{6}, 7298 (2015).

\bibitem{Ando} T. Ando, A. B. Fowler and F. Stern, \rmp~\textbf{54} 437
(1982).

\bibitem{Beenakker} C. W. J. Beenakker and H. van Houten, in \textit{Solid
State Physics}, \textbf{44}, (H. Ehrenreich and D. Turnbull, eds., Academic
Press, Boston), pp. 1-228 (1991).

\bibitem{Z1} N.~V.\ Zavaritskii and Z.~D. Kwon, Pis'ma Zh. Eksp. Teor. Fiz. 
\textbf{39},61 (1984) [JETP Lett., \textbf{39}, 71 (1984)].

\bibitem{Z2} N.~V. Zavaritskii and I.~M. Suslov, Zh. Eksp. Teor. Fiz., 
\textbf{87}, 2152 (1984) [Sov. Phys. JETP, \textbf{60}, 1243 (1984)].

\bibitem{BPV92} Ya. M. Blanter, A.~V.~Pantsulaya, A.~A. Varlamov, Phys. Rev.
B \textbf{45}, 6267 (1992).

\bibitem{Bob71} B.~I. Ivlev, G.~M. Eliashberg, Pis'ma Zh. Eksp. Teor. Fiz. 
\textbf{13}, 464 (1984) [JETP Lett., \textbf{13}, 333 (1971)].

\bibitem{Abrikosov} A.~A. Abrikosov \emph{Fundamentals of the Theory of
Metals, } Elsevier, Amsterdam (1989),\S 19.5, page 414.

\bibitem{Dutton1958} D. Dutton, Phys. Rev. \textbf{112}, 785 (1958).

\bibitem{Mahan} G. D. Mahan, ``Many-Particle Physics'' (Kluwer
Academic/Plenum Publishers -- New York, Boston, Dodrecht, London, Moscow),
p. 380 (1981).

\bibitem{Novikov} D.~S. Novikov, Phys. Rev. B \textbf{79}, 235304 (2009).
\end{thebibliography}
\end{document}